# Evolutionary Search and Theoretical Study of Silicene Grain Boundaries' Mechanical Properties

Jianan Zhang (1), Aditya Koneru (1, 2), Subramanian K. R. S. Sankaranarayanan* (1, 2), & Carmen M. Lilley* (1)

(1) Department of Mechanical and Industrial Engineering, The University of Illinois at Chicago, 842 W. Taylor, Chicago, Illinois 60607, USA.

(2) Center for Nanoscale Materials, Argonne National Lab, Argonne, Illinois 60439, USA.

Emails:

Subramanian K. R. S. Sankaranarayanan: skrssank@uic.edu; ssankaranarayanan@anl.gov

Carmen M. Lilley: clilley@uic.edu

**Keywords:** Silicene, Grain Boundary, Mechanical Property, Genetic Algorithm, First Principle Study, 2D Materials

## Abstract

Defects such as grain boundaries (GBs) are almost inevitable during the synthesis process of 2D materials. To take advantage of the fascinating properties of 2D materials, understanding the nature and impact of various GB structures on the pristine 2D sheet is crucial. In this work, using an evolutionary algorithm search, we predict a wide variety of silicene GB structures with very different atomic structures compared to those found in graphene or hexagonal boron-nitride. Twenty-one GBs with the lowest energy were validated by density functional theory (DFT) - a majority of which were previously unreported to our best knowledge. Based on the diversity of the GB predictions, we found that the formation energy and mechanical properties can be dramatically altered by adatoms positions within a GB and certain types of atomic structures, such as four-atom rings. To study the mechanical behavior of these GBs, we apply strain to the GB structures stepwise and use DFT calculations to investigate the mechanical properties of 9 representative structures. It is observed that GB structures based on pentagon-heptagon pairs are likely to have similar or higher in-plane stiffness and strength compared with the zigzag orientation of pristine silicene. However, an adatom located at the hollow site of a heptagon ring can significantly deteriorate the mechanical strength. For all the structures, the in-plane stiffness and strength were found to decrease with increasing formation energy. For the failure behavior of GB structures, it was found that GB structures based on pentagon-heptagon pairs have failure behavior similar to graphene. We also found that the GB structures with atoms positioned outside of the 2D plane tend to experience phase transitions before failure. Utilizing the evolutionary algorithm, we locate diverse silicene GBs and obtain useful information for their mechanical properties.

## 1. Introduction

The discovery of Graphene has propelled investigations aimed at the discovery and design of other possible analogous two-dimensional 2D materials (1-4). Among these 2D materials, silicene, formed by carbon's neighbor on the periodic table, has been theoretically shown to form a buckled honeycomb lattice while preserving a similar Dirac cone electronic structure as graphene (5-8). The fabrication of silicene was achieved by Vogt *et al*. (9), who successfully synthesized large-scale silicene on a single crystal Ag substrate. Although pristine silicene, just like graphene, cannot be directly used for building semiconductor devices due to its zero band gap (10), previous researchers have demonstrated that the band gap for monolayer silicene is controllable via an external electric field, in contrast to graphene (11). This potential for tunable electronic properties makes it a promising material to study for emerging electronic devices.

In addition to the limitations of the band gap size, the mechanical properties of silicene are also crucial for the implementation of possible nanodevices. Yet, much of the existing computational research has focused on the mechanical properties of pristine silicene (12-16), such as in-plane stiffness and the ultimate strength. Since defects are almost inevitable during the synthesis process (17), it is also necessary to look into changes in the mechanical properties introduced by various defects. In the work of Liu et al. (18), when biaxial tensile tests were applied to polycrystalline silicene sheets, the fracture strength decreased while the fracture strain increased with decreasing grain size. The decrement of the fracture strength was attributed partly to the increasing area with defects in the GB. In the work of Rakib et al. (19), it was reported that a switch from an inverse pseudo Hall-Petch to a pseudo Hall-Petch behavior was observed at the critical grain size of 17.32 nm for nanocrystalline silicene. These studies investigated the correlation between GB size and the strength of the polycrystalline silicene. However, the defect types or their orientation with respect to the applied strain to the silicene sheet were not studied. Other existing work on silicene point defects and GBs (20-23) demonstrated that with different types of defects, different mechanical properties and electronic properties are possible. However, despite the selected types of defects studied in the abovementioned work, the GB atomic structure in these studies was limited to the interface between predetermined fixed orientations of the pristine sheets on two sides (e.g. the interface between sheets with armchair and zigzag orientations) or introduced by reconstruction of a few types of point defects (e.g. single and double vacancy). Previously work by Wang (24) discovered that silicene exhibits $sp^3$-like behavior due to partial hybridization between $s$ and $p^z$ orbitals. We hypothesize that unlike $sp^2$ hybridization in graphene, this $sp^3$-like behavior allows for much more diverse GB structures in silicene, and thus the possible GB structures for silicene are far from being thoroughly explored. As a result, the mechanical properties of grain boundaries of silicene have also not been fully explored. By utilizing an evolutionary searching technique and DFT calculations, we can address a broader understanding of the formation energy and mechanical property impact from GB defects in silicene.

Genetic algorithms (GA) as an evolutionary approach for the optimization problem have been extensively used for material design problems, such as interface or GB structure predictions and crystal structure searches (25-32). In research (33), a workflow based on a GA was introduced for predicting energetically favorable GB structures for 2D materials. In our recent work, the GA method was benchmarked with graphene, and 128 new GB structures were predicted and analyzed in detail. Our statistical analysis of the formation energies for these structures revealed lateral interface predictions consistent with research on graphene defects (34).

In this work, we build upon the GA workflow by [32], as shown in Fig. 1, by improving the GA search with a graph isomorphism check to predict multiple metastable phases for fixed combinations of laterally interfacing silicene sheets. A total of 5000 generations of GA searching were performed for each combination of the pristine silicene domains, and for each combination, multiply distinct silicene GB structures were predicted. From this pool of structures, 21 GB systems were validated by DFT simulations. Upon relaxation, 17 previously unreported GB structures were found for silicene. These validated structures were then studied for the GB material properties, and we found that the GBs were composed of various types of defects including SW defects, adatoms, four-atom, and five-atom rings. Although the GBs based on SW defects were generally found to have lower formation energy, which is similar to graphene, we found that a Si adatom may increase or dramatically decrease the formation energy depending on where the adatom located. It is also observed that four-atom rings, which is uncommon in graphene (35) and hexagonal boron-nitride (h-BN) (36), can lower the formation energy of the GB by reducing strain.

With a well-sampled population of GB structures, we classified all the structures into different types based on their topological features. Nine representative structure was selected, and stepwise uniaxial strain was applied until failure of the structure occurred. The stress-strain curve for all GBs and two orientations of pristine sheet were plotted. The mechanical properties such as the in-plane stiffness and in-plane strength was evaluated and compared to similar GB structures of other 2D materials. We found that the majority of GBs with SW defects for silicene have in-plane stiffness and in-plane strength comparable or higher than the zigzag direction of the pristine sheet. In addition, the fracture behavior of these GB structures are similar to those of graphene, where the fracture initiates from the bond shared by a heptagon ring and a neighboring hexagon ring (35). However, a GB with an adatom located at the hollow site of a heptagon is an exception which exhibits significantly lower stiffness, strength and very different failure behavior. We attribute the poor stiffness and strength to a prestrain for the bonds introduced by the adatom. Interestingly, the adatom located at top site or hollow site of hexagon does not impact the mechanical behavior. Finally, unlike the brittle fracture behavior of h-BN (36, 37), many silicene GBs predicted in our work change to other topological structures when strain is applied. For example, out-of-plane atoms move to positions within the silicene plane as the sheet is strained and leads to sudden change in the gradient of the strain-stress curve.

In summary, by utilizing a powerful evolutionary search method, previously unreported silicene GBs were located. A comprehensive analysis based on first-principle method was performed to study the formation energy and mechanical properties of these GBs. The theoretical results provided useful insight to the effects GBs have on mechanical behavior of silicene under uniaxial tensile strain. The workflow in this work combining GA search and DFT simulations, as demonstrated in Fig. 1, is anticipated to accelerate the discovery of 2D interfaces and GBs.

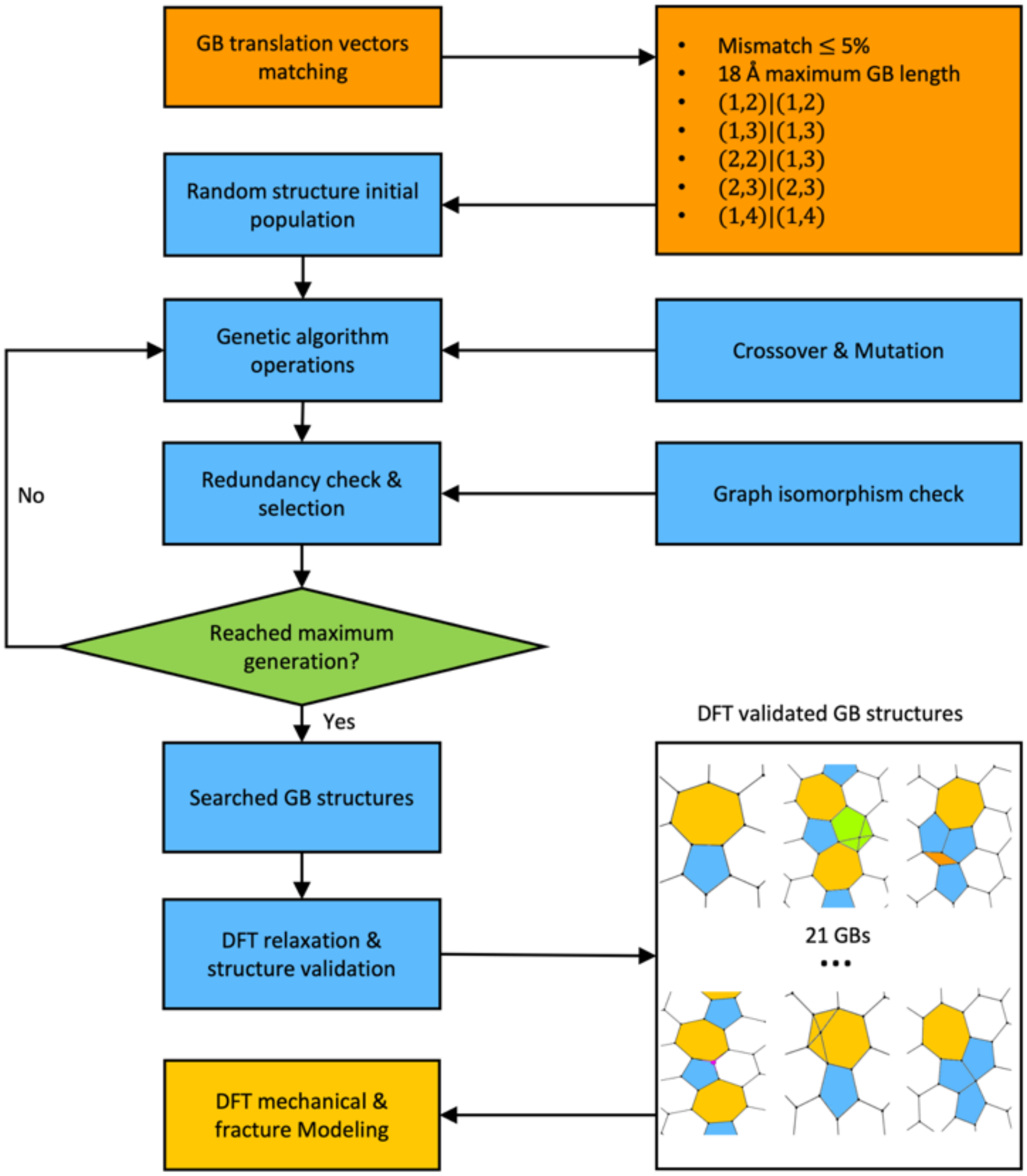

*Figure 1: Flowchart of the workflow implemented in this paper. Start from the structure search, DFT relaxation and validation, then followed by first-principle study of the silicene grain boundaries' properties.*

# 2. Methods

## 2.1. Structural Model and GA Search

The translation vectors $(m_l, n_l)$ and $(m_r, n_r)$ for the left and right bulk silicene sheets, adjacent to the GBs, are first determined to establish the rotation angles of the bulk sheets (33, 35, 38-40). Herein, we label GB structures with different translation vectors as $(m_l, n_l)|(m_r, n_r)$-$iX$ where $(m_l, n_l)$ and $(m_r, n_r)$ are the translation vectors of the left and right side, respectively. The notation $iX$ ($X = 1,2,3$ ...) is the index to distinguish different structures with the same translation vector pair. The length of the supercell along the GB direction is then calculated for the left and right domains with the following equations (33, 35, 38, 40):

$$L_l = a_0\sqrt{n_l^2 + n_l m_l + m_l^2} \quad (1)$$

$$L_r = a_0\sqrt{n_r^2 + n_r m_r + m_r^2} \quad (2)$$

The variable $a_0$ is the length of the unit vector for silicene and was set to 3.87Å based on our first-principles calculation. This value is comparable to values reported in existing research (15, 41, 42). The commensurateness for $L_l$ and $L_r$ determines whether realistic GBs exist for such a combination. As mentioned in the work done by Zhang *et al.* (35), the mismatch between $L_l$ and $L_r$ increases formation energy by introducing strain energy. Therefore, in the case where $L_l$ and $L_r$ are different (rotation angles for the two domains are not identical), we adopted the same 5% mismatch cutoff as Zhang *et al.* and filtered out the combinations with a greater mismatch between the translation vectors. This mismatch cutoff value is selected based on the previous research work(43) showing that in the context of strain engineering of 2D materials, the most significant property changes occur when less than 2.5% strain is applied. Therefore, we believe that 5% cutoff is a safe limitation without excluding potentially promising grain boundary (GB) structures. This step makes sure that only the physically realistic GBs are considered. The larger the components of the lattice vectors, the larger the final supercell sizes will be, thus increasing the cost for DFT simulations. Therefore, we also limited the length of a GB to a maximum length of 18 Å. With these restrictions, the GB structures for the following 5 combinations of lattice vector components: (1,2)|(1,2), (1,3)|(1,3), (2,2)|(1,3), (2,3)|(2,3) and (1,4)|(1,4) were studied. The rotation angle between the pristine silicene and the corresponding left and right bulk domain is $\theta_l$ and $\theta_r$, which can be calculated using the following equations:

$$\theta_l = tan^{-1}\left(\frac{\sqrt{3}m_l}{(m_l + 2n_l)}\right) \quad (3)$$

$$\theta_r = tan^{-1}\left(\frac{\sqrt{3}m_r}{(m_r + 2n_r)}\right) \quad (4)$$

The misorientation angle for a GB structure is defined as $\theta = \theta_l + \theta_r$.

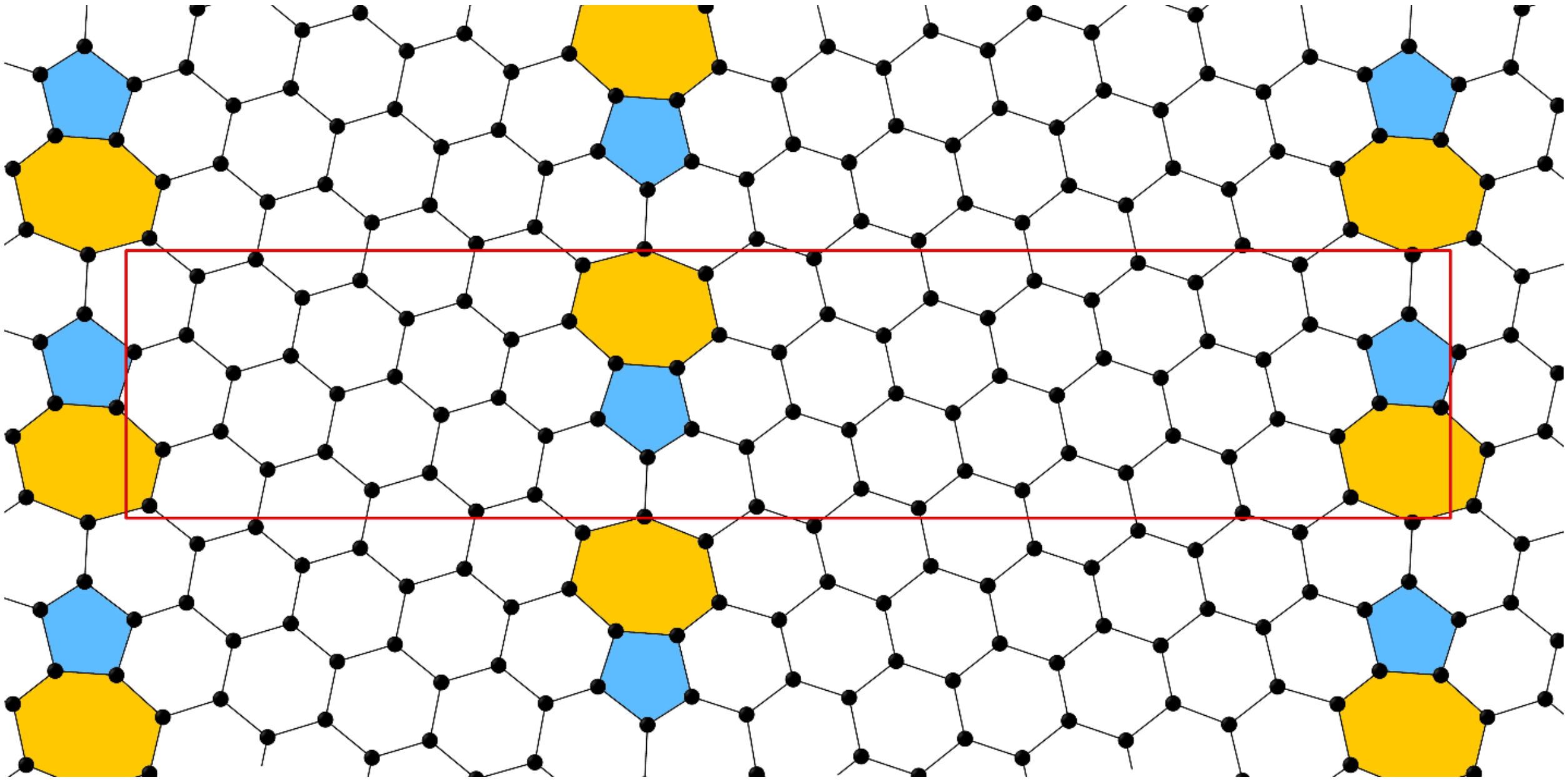

*Figure 2: Supercell for grain boundary structure (1, 2)|(1, 2)-i1 predicted by genetic algorithm. The red lines are the boundary of the simulation box.*

In a departure from the GA search implemented in Reference (33), we used a supercell containing two instead of one GB as shown in Fig. 2 so that two GBs are in the same supercell and a periodic boundary condition is enforced within the 2D sheet plane. This supercell structure allows us to measure the formation energy of each structure more accurately. In addition to the new type of supercell, we also introduced the graph isomorphism check as implemented in the NetworkX (44) package for comparing the connectivity of nodes between graphs (GB structures) and filtering out duplicate GBs. Finally, the GBs of 2D materials can be naturally transferred to a mathematical graph, where one can treat the atomic positions as the nodes and the interatomic distances as the edges. After we transfer structures into graphs, the isomorphism check considers whether two structures are identical if the topological structure is the same. The structures with the same type of defects (e.g., Stone-Wales defect) but with only slightly different bond lengths will be detected and discarded. In this way, multiple metastable phases of GBs for each combination of lattice vectors can be predicted during the search.

## 2.2. Empirical Potential Minimization Details

During the GA search, geometry optimization using an in-house Tersoff potential (45) was integrated with Large-scale Atomic/Molecular Massively Parallel Simulator (LAMMPS) version 27 May 2021 (46). For the minimization criterion, the forces on all the atoms were relaxed until they were smaller than $10^{-5}$ ${}^{eV}/_{Å}$ or a maximum of $10^{4}$ evaluations of energy or forces were completed. This local optimization of the GB structures allows the GA search for a global minimum to converge more quickly and avoids the searching process getting trapped to a local minimum.

### 2.3. DFT Simulation Details

The Vienna ab initio simulation package 5.4.1 (VASP 5.4.1) (47) is used for the DFT study of material properties and geometry optimization. Projector augmented wave pseudopotential was used and the Perdew Burke Ernzerhof (PBE) functional was implemented for the treatment of electron exchange and correction. The kinetic energy cutoff was set to 500 $eV$, the vacuum space between neighboring images of the 2D sheet was set to 16 Å, and the two GBs within each supercell were separated by pristine silicene with a width equal to 18 Å. The K-point spacing was set to 0.035 $Å^{-1}$ in reciprocal space to ensure the accuracy of simulations for different supercells with different lengths and widths of GBs. For the geometry relaxation of the GB structures, the energy and force criteria are set to $10^{-6}$ $eV$ and 0.03 $eV/Å$, respectively.

### 2.4. Evaluation of mechanical properties

We studied the stress-strain response of the silicene GBs obtained from the GA search. Strain, defined as $\epsilon = \frac{a-a_0}{a_0}$, was applied to the supercells stepwise and perpendicular to the GB. The variable $a$ is the strained supercell length perpendicular to the GB and $a_0$ is the original supercell length. A series of 1% increments of strains were applied to the structures. At each strain step, the atomic positions were relaxed without constraint using the conjugate gradient algorithm as implemented in VASP, and the cell dimension perpendicular to the strain direction was allowed to relax, while the cell dimension along other directions was kept fixed. The relaxed structure is subsequently strained at the next step to guarantee the continuation of the strain steps in the modeling. We adopt the same material characteristic equations previously introduced by of Zhao *et al.* (14) for silicene. In-plane stiffness, $C$, and in-plane stress, $f$, are used to quantify the stiffness and stresses within the sheet, and both $C$ and $f$ have units of $N/m$. For the simulation supercell of each GB structure, vacuum layers were added perpendicular to the 2D sheets to avoid the influence of neighboring images. The DFT evaluated stress was computed for the entire supercell. Therefore, to obtain the in-plane stress for the 2D sheet, the DFT calculated stress value was rescaled by the height of the supercell for each strain step to obtain the value. The in-plane stiffness is calculated by applying a linear regression using the in-plane stress, $f$, curves for strains less than 5%. The GB structures were strained until $f$ dramatically dropped, which indicates that the structures were either fractured (35) or only an atomic chain remained between two domains. We consider both scenarios as the failure of the silicene structure.

### 2.5. Code and software version

Our GA code was implemented using Python 3.7 along with packages: Distributed Evolutionary Algorithms in Python v1.3.3(48), Atomic Simulation Environment v3.22.1(49).

## 3. Results

### 3.1. A sampling of Silicene GB's

GA searching and sampling of the GB structures were performed for the lattice vector combinations discussed in the Methods section. The ten structures with the lowest formation energy from 5000 generations of searching were recorded for each orientation. After duplicate structures were removed, the remaining 31 GB structures were simulated using DFT to perform structural relaxations. If a structure did not converge to the force limit, as mentioned in the method section, or the structure reconstructed to an existing GB structure in the population being relaxed, the structure was removed. The formation energy of the GB structures was evaluated with the following equation (33, 35, 38, 40):

$$E_f = \frac{(E_t - N \times e)}{2L} \quad (5)$$

where, $E_f$ and $E_t$ are the formation energy of the GB and the total energy of the structure, respectively. $N$ is the number of atoms and $e$ is the energy per atom for pristine silicene. The value of $e$ was set to -4.79 eV based on our first-principle calculation of pristine silicene. $L$ in Eq. 5 represents the length of the GB, the factor 2 in the denominator accounts for two identical GBs in one supercell.

Eventually, a total of 21 fully relaxed GB structures from 5 different combinations were obtained, 17 structures of which were previously unreported to our best knowledge. For the 21 GB structure predicted, atomic structures that are infrequently seen for graphene or h-BN (e.g., hollow or top site adatoms and a four-atom ring) were found in many of structures. Previous work done by Wang (24) revealed that 2D silicene sheet shows $sp^3$-like behavior due to partial hybridization between the $s$ and $p_z$ orbital. It is believed that this $sp^3$-like hybridization for silicene sheet leads to not only the buckled 2D sheet but also the diverse GB structures, which is very different from graphene with $sp^2$- and $\pi$-bonds.

To study the effects these diverse GBs have on mechanical properties and the formation energy, we classified the 21 predicted GBs into 6 types described as follows: (1) SW defects that are periodically separated by hexagon rings along the length of the GB region, as shown in Fig. 3. (2) Periodically repeating SW defects along the length of the GB region, shown in Fig. 4(a-c). (3) SW defects with extra pentagons along the length of the GB region, shown in Fig. 4(d-f). (4) 5 and 7-atom rings that are separated by hexagons or 4-atoms along the length of the GB region, shown in Fig. 5(a-c). (5) Multiple 5 and 7-atom rings clustered along the length of the GB region with 7-atom rings neighboring each other, as shown in Fig. 5(d-f). (6) GB regions with an octagon, shown in Fig. 5(g). Pentagons are colored in blue, heptagons are in yellow, four-atom rings are in orange, a hexagon with an adatom at the hollow site is labeled in green, eight-atom rings are in red, and both the adatom on the top side and the triangles in the GB structures are in purple. GBs that have additional topological features (e.g., adatoms, 4-atom rings) are also included in the corresponding classes based on the above-mentioned topological features.

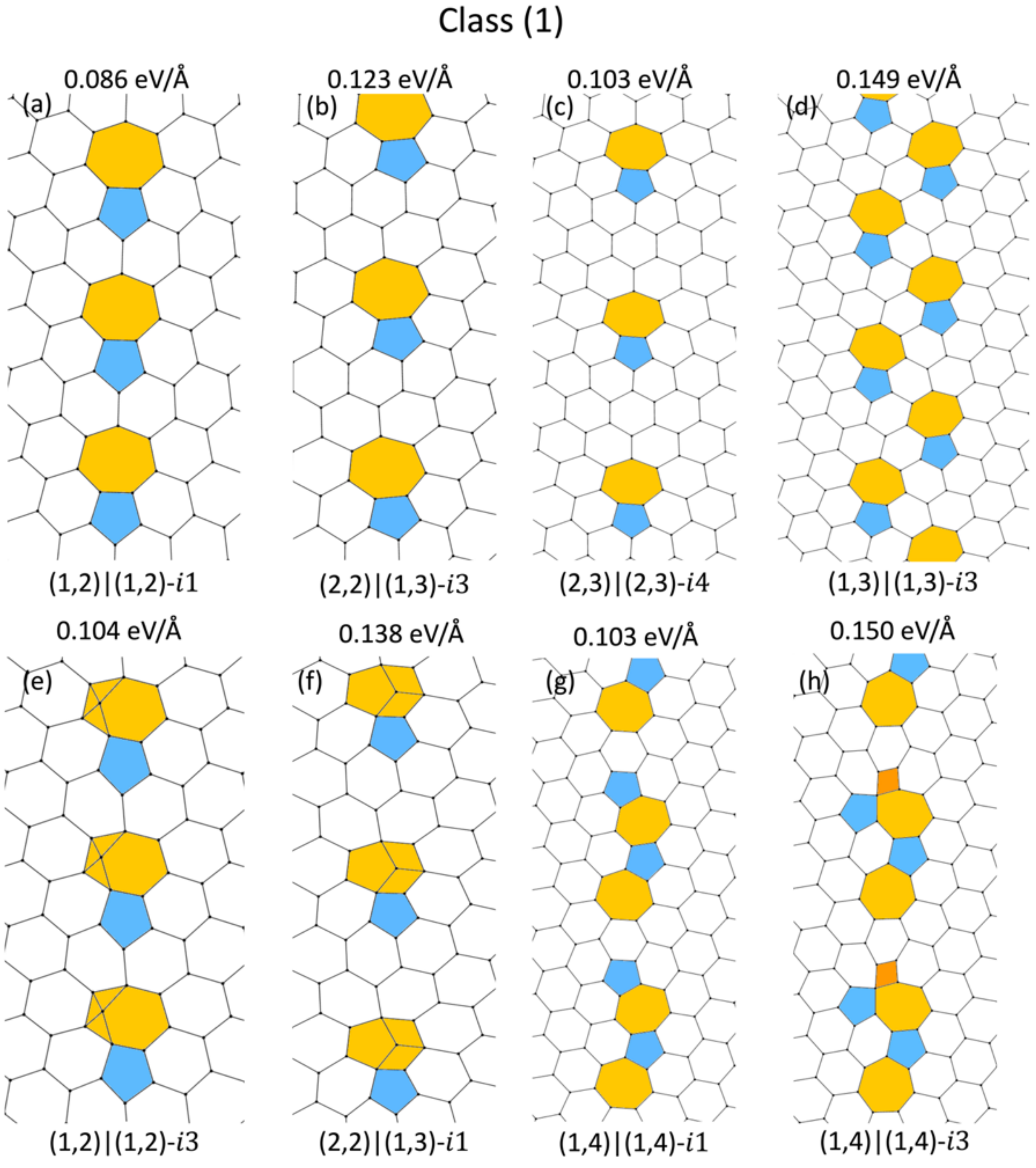

*Figure 3: Grain boundary structures in class (1). All structures are based on Stone-Wales defects that are periodically separated by hexagons fill the Grain boundary region.*

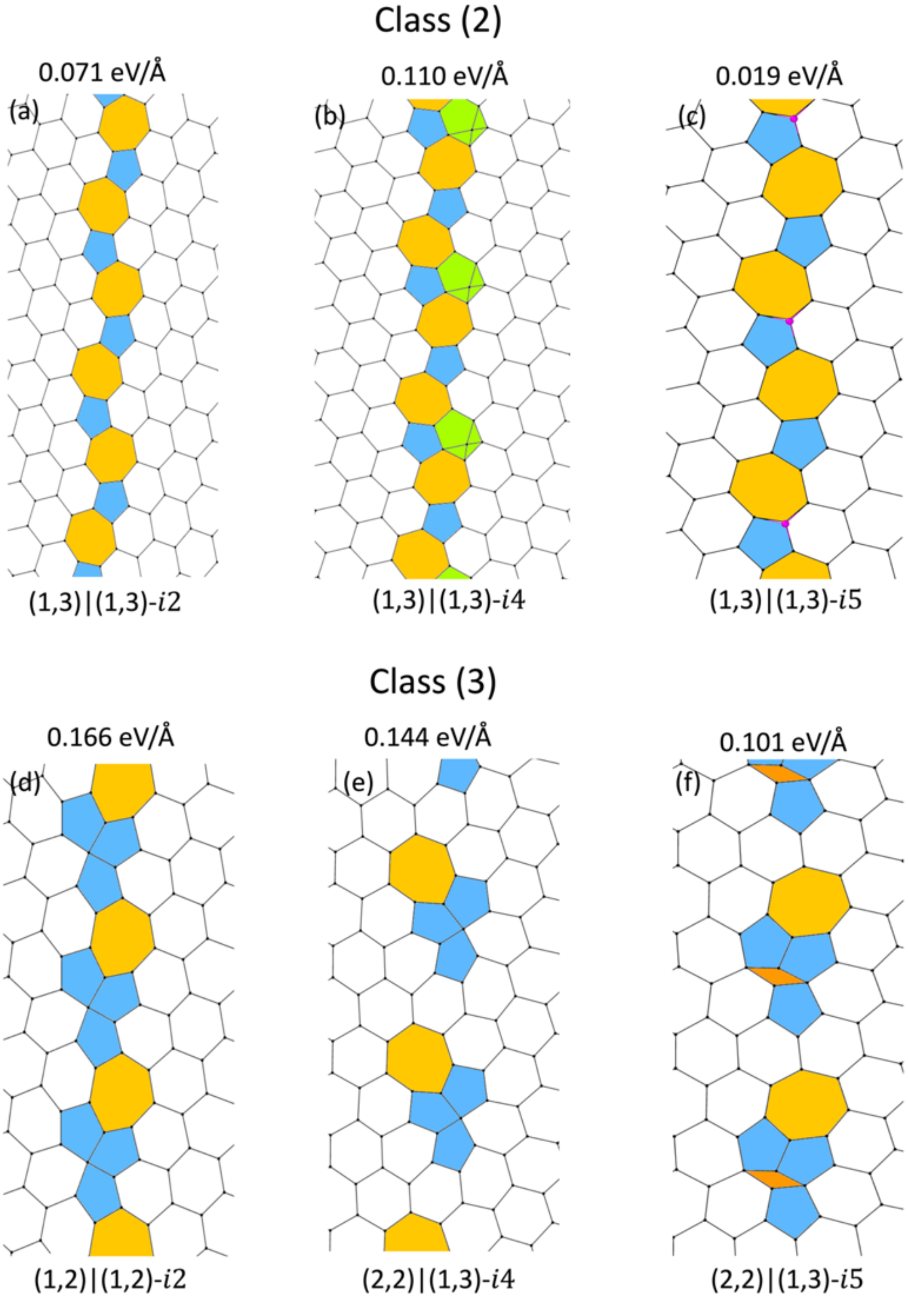

*Figure 4: Structures in class (2) & (3). For class (2) periodically repeating Stone-Wales defects along length of the grain boundary region. For class (3) Stone-Wales defects with extra pentagons filled the grain boundary region.*

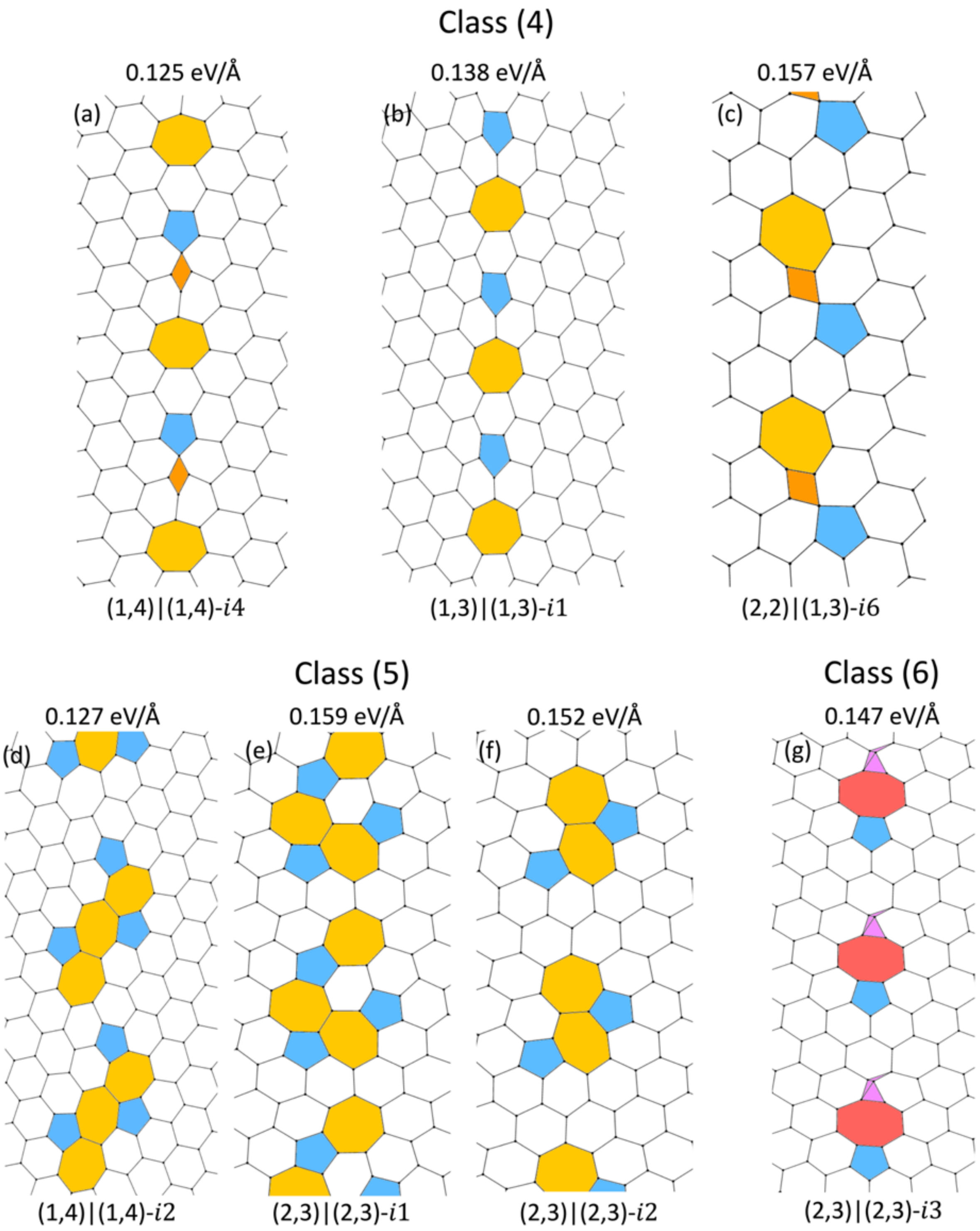

*Figure 5: Structures in class (4), (5) & (6). Structures in class (4) have 5 and 7-atom rings that are separated by hexagons or 4-atom rings filled the GB region. Structures in class (5) have clustered 5, 7-atom rings with 7-atom rings connected. Class (6) structures have octagons located in the grain boundary region.*

The formation energy for all the GB structures obtained range from 0.019 to 0.419 eV/Å, which is much lower than previously reported values for graphene (from 0.28 to 0.8 eV/Å) (40), h-BN (from 0.26 to 0.91 eV/Å)(36). Silicene GBs also had structures with much lower formation energy values, when compared with previously reported phosphorene GB values (from 0.09 to 0.24 eV/Å) (38). Previous studies for graphene (35) and h-BN (36) found that some GB structures induce a large out-of-plane inflection angle, and the intrinsic strength of these GB structures decreases as the inflection angle increases. In contrast, previous research for 2D black phosphorene (38, 50) found that GB structures tend to remain flat or show very small out-of-plane buckling due to the flexible tetrahedral geometry formed by the $sp^3$ P-P bonds. For the Silicene GB structures in this work, no large inflection angles were observed and only some wrinkling for (2, 3)|(2, 3)-$i3$, (1, 3)|(1, 3)-$i1$, (2, 2)|(1, 3)-$i3$ and (2, 3)|(2, 3)-$i4$. We also found that some silicene GBs have out-of-plane atoms at the GB regions. These structures have atoms at GB region that have Z coordinates at least 1 Å higher or lower than the middle height of the supercell where the bulk silicene was placed. We observe no correlation between the presence of these out-of-plane atoms and the wrinkling in the structures. As mentioned above, previous work done by Wang (24) found that the monolayer Silicene tends to show $sp^3$-like behavior. We believe that the partial hybridization of $s$ and $p_z$ bonds in Silicene is the reason GBs have little to no inflection angles. We also highlight the previous work on Silicene defects studied using molecular mechanics and molecular dynamics methods (23), where it was reported that the GB structures with 5-atom and 7-atom paired rings can have inflection angle as large as 30 degrees and the intrinsic strength decreased when inflection angle increase. Our results were not consistent with this study possibly due to the difference in methodology such as different force field and DFT. The formation energy of the GB structures in this study and the studies mentioned above is summarized in Table 1 below.

*Table 1: Class and formation energy of grain boundaries on silicene and other two-dimensional materials from literature. Structures with out-of-plane atoms as defined above is labeled by asterisk in table. Adjacent rows shaded in same color are GBs on different 2D materials but with similar atomic structures for comparison.*

| Class | GB | $E_f$ (eV/Å) |
|---|---|---|
| 1 | (1,2)\|(1,2)-i1 | 0.086 |
| | (1,2)\|(1,2)-i1 (graphene) | 0.34 (35), 0.344 (40) |
| | (1,2)\|(1,2)-i1 (h-BN) | 0.51 (36) |
| | (2,2)\|(1,3)-i3* | 0.123 |
| | (2,2)\|(1,3)-i3 (graphene) | 0.51 (35) |
| | (2,3)\|(2,3)-i4* | 0.103 |
| | (2,3)\|(2,3)-i4 (graphene) | 0.36 (35), 0.351 (40) |
| | (1,3)\|(1,3)-i3 | 0.149 |
| | (1,2)\|(1,2)-i3* | 0.104 |
| | (2,2)\|(1,3)-i1* | 0.138 |
| | (1,4)\|(1,4)-i1* | 0.103 |
| | (1,4)\|(1,4)-i1 (graphene) | 0.41 (35), 0.399 (40) |
| | (1,4)\|(1,4)-i3* | 0.150 |
| 2 | (1,3)\|(1,3)-i2 | 0.071 |
| | (1,3)\|(1,3)-i2 (graphene) | 0.33 (35), 0.291 (40) |
| | (1,3)\|(1,3)-i4* | 0.110 |
| | (1,3)\|(1,3)-i5* | 0.019 |
| 3 | (1,2)\|(1,2)-i2* | 0.166 |
| | (2,2)\|(1,3)-i4* | 0.144 |
| | (2,2)\|(1,3)-i5* | 0.101 |
| 4 | (1,4)\|(1,4)-i4* | 0.125 |
| | (1,3)\|(1,3)-i1* | 0.138 |
| | (1,3)\|(1,3)-i1 (graphene) | 0.503 (40) |
| | (2,2)\|(1,3)-i6* | 0.157 |
| 5 | (1,4)\|(1,4)-i2 | 0.127 |
| | (2,3)\|(2,3)-i1 | 0.159 |
| | (2,3)\|(2,3)-i2 | 0.152 |
| 6 | (2,3)\|(2,3)-i3* | 0.147 |

The formation energy of GBs in 2D materials composed solely of one type of element arises from two main sources: the energy associated with mismatch strain and the alterations in atomic structure compared to a pristine sheet. For example, structures (1, 2)|(1, 2)-$i1$ and (2, 2)|(1, 3)-$i3$ shown Fig. 3 (a) and (b), respectively, have the same SW defects. However, the energy for the (1, 2)|(1, 2)-$i1$ structure is lower due to the length of the left and right translation vectors being identical, and therefore there is no strain energy introduced by mismatch of two domains as compared to the (2, 2)|(1, 3)-$i3$ structure that has a mismatch.

We also noticed the presence of an adatom located at different locations in the GB regions have different effects on the formation energy. For example, for structures shown in Fig. 3(a) and (e) both GB structures have 5-atom and 7-atom ring pairs. Yet, when an adatom is located at a hollow site, Fig. 3(e), the formation energy of Fig. 3(e) increases by 0.018 eV/Å. This indicates that the adatom located at the hollow site of the heptagon is energetically unfavorable and was further confirmed by comparing the formation energy between Fig. 3(b) and 3(f), where the formation energy increases by 0.015 eV/Å for Fig. 3(f) due to the hollow site adatom within a heptagon. Class (2) defects, as shown in Fig. 4(a) and (b) (1, 3)|(1, 3)-$i2$ and (1, 3)|(1, 3)-$i4$, have periodically repeating SW defects along the GB length. However, the structure in Fig. 4(b) has a formation energy that is 0.039 eV/ Å higher due to an adatom located at the hollow site of the hexagon next to the periodic pentagon-heptagon pairs. It is believed that although adatoms at a hollow site of hexagon or heptagon rings are energetically unfavorable, the adatom at the hexagon hollow site introduces more strain due to the smaller area of the hexagon.

It is also worth noting that even though hollow site adatoms lead to higher formation energy, for the one GB structure predicted with top-site adatom the formation energy is significantly lowered. By comparing the formation energy between two distinct GB structures with the same (1, 3)|(1, 3) interface, as seen in Fig. 4(a) and Fig. 4(c), it is noted that an adatom on the top site of a Si atom will significantly reduce the formation energy from 0.071 eV/ Å to 0.019 eV/ Å. As reported in the review by Zhao *et al.* (10), the top-site atom will vertically deflect the bottom silicon atom so they form a dumbbell like configuration. This dumbbell like configuration forms a local $sp^3$ hybridization which leads to a significant drop of the formation energy. Previous research (22, 51) also found similar top-site adatom point defect structures have negative formation energy. GBs with top-sites adatoms are more probable due to the much lower formation energies compared with other defects.

One would expect that the formation energy increases as more non-hexagonal rings form in the GB region, however we noticed that for silicene GBs, the presence of four-atom ring would lower the formation energy and the presence of additional 5-atom ring at the GB region would increase the formation energy. For example, the (1, 2)|(1, 2)- $i2$ GB in Fig. 4(d) is formed by pentagon-heptagon pairs, with extra pentagons and higher formation energy comparing to the (1, 2)|(1, 2)-$i1$ structure in Fig. 3(a). When comparing the (2, 2)|(1, 3) GB in Fig. 4(e) to the (2, 2)|(1, 3) GB in Fig. 4(f), it is noticed that the four-atom rings between non-SW pentagons can lower the formation energy of the GB.

In Fig. 6, we plot the distribution of the formation energy of all the GB structures with respect to their classes. In addition to the atomic features that are used to classify the structures, other atomic features for each of the structures is also labeled as in the legend. The structures with out-of-plane atoms at the GB regions are circled in blue. From the plot it is obvious that the formation energy for structures that have class (1) and class (2) features are lower on average than the other classes, the structure (2, 3)|(2, 3)-$i3$ with 8-atom ring and multiple 3-atom rings in class (6) has formation energy higher than the average formation energy for other classes, indicating the atomic structure for the GB is energetically unfavorable. As can be noted, the average formation energy for structures in class (5) is comparable to the formation of the structure in class (6). In addition, two out of three structures in class 5 have higher formation energies as compared with the structure in class 6. However, the defects in class (5) GBs are mainly formed by 5, 7-atom rings similar to the GBs in class (1) and (2), which have a lower average formation energy. As such, this indicates that the higher formation energy of structures in class (5) is likely attributed to the higher concentration of 5, 7-atom rings, i.e. the presence of more defects, in the GB. For GBs from class (6), the higher formation energy comparing with GBs from other classes is introduced by different types of atomic structures, such as the 5-atom ring, 8-atom ring, and the 3-atom rings in the structure. Most of the structures have out-of-plane atoms in the GB region, while the structures that belong to Class (5) with 7-atom rings directly touching each other do not have out-of-plane adatoms. No correlation between out-of-plane atoms and the formation energy is observed. 5 combinations of lattice vector components as discussed in the Method section have misorientation angle 21.8, 27.8, 38.2, 43.9 and 46.8 degree, and no obvious correlation was observed between the formation energy and these misorientation angles.

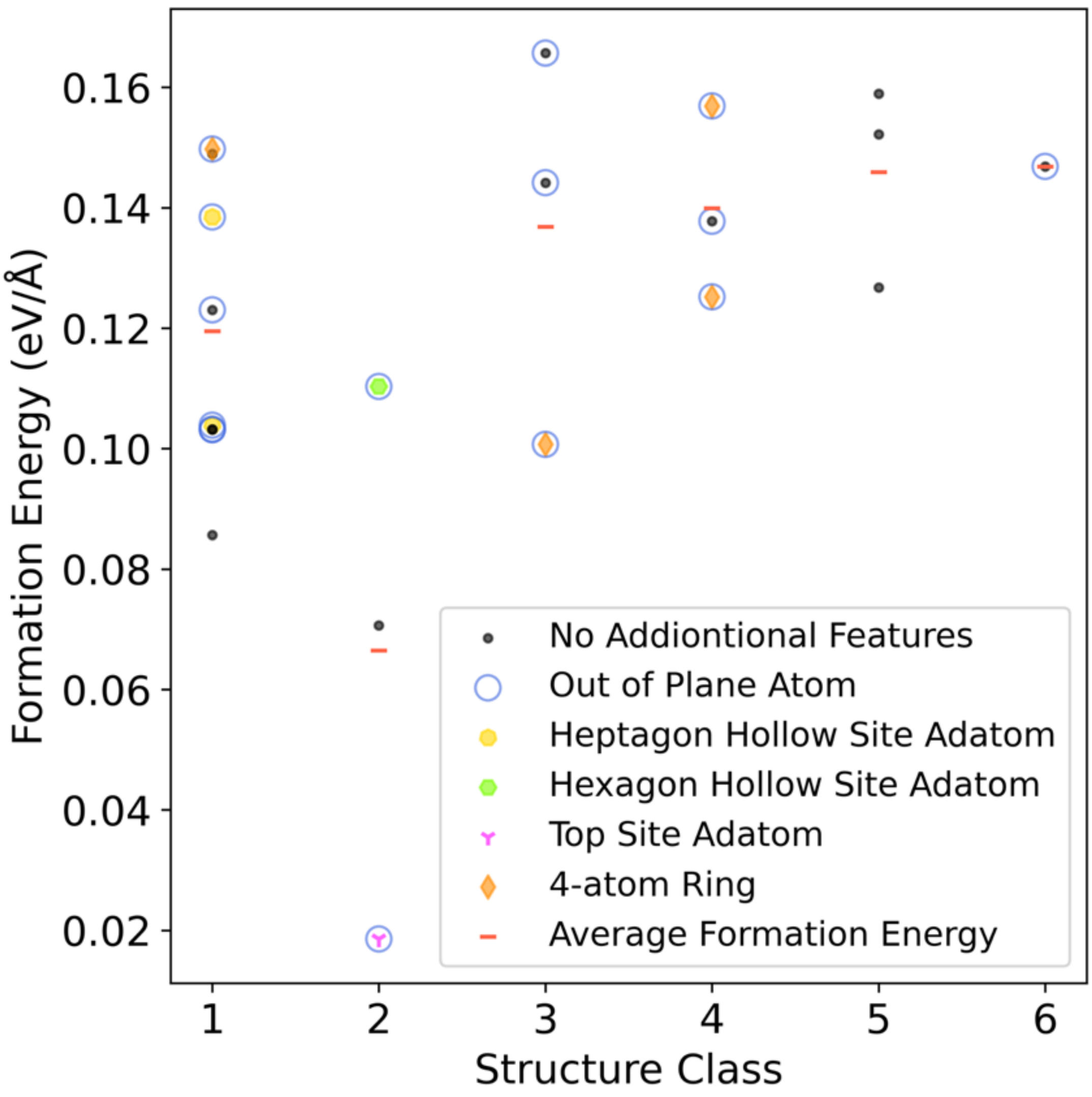

*Figure 6: Formation energy of the structures based on their classes. The structures with additional features, including adatom at various of locations, 4-atom ring along with the average formation energy of each structure class are labeled as in legends. The structure with out of plane atoms at GB regions is labeled with blue circles.*

### 3.2. In-plane Stiffness and In-plane Strength

After the relaxed structures were obtained through the optimization process, we evaluated the influence of GB defects on the silicene mechanical properties, specifically the stress-strain response (see Methods for details). Considering the high cost of DFT simulations, and to avoid the study of repetitive structures, GB structures with representative defect features such as 5, 7-atoms rings, multiple neighboring 5-atom rings, hollow and top site adatoms, 4-atom rings and 8-atom rings with 3-atom rings were selected and strained as outlined in the Methods section. No structure in class (5) was selected as all GBs in the class (5) are formed by the 5, 7-atom ring pairs which is similar to the GBs in class (1) and (2) in terms of the atomic structures. We focused on the structures with special atomic structures feature for better exploration of the impact on the mechanical properties introduced by these atomic structure features. The in-plane stress, $f$, for each strain step is plotted in Fig. 7. The curves for armchair orientation, labeled with blue circles, and zigzag orientation, labeled with orange triangles, of pristine silicene are also plotted for comparison.

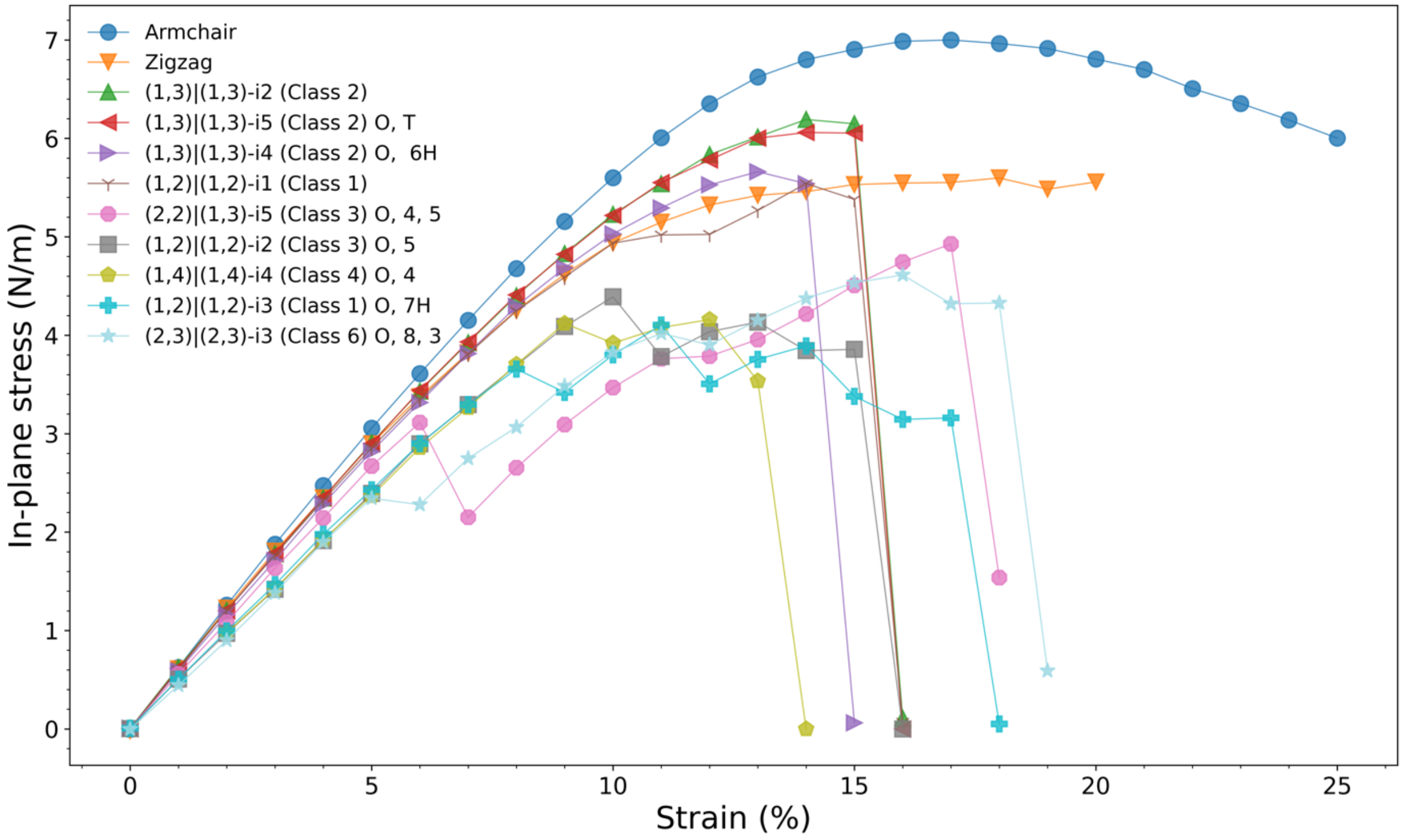

*Figure 7: In-plane stress versus strain for all selected grain boundary structures and pristine silicene in the armchair and zigzag direction. Different atomic structure features of GBs are labeled based on the following notation in the legend: O: Out-of-plane atom; T: Top site adatom; 6H: Hexagon hollow site adatom; 7H: 7-atom ring hollow site adatom; 3: 3-atom ring; 4: 4-atom ring; 5: Multiple 5-atom ring; 8: 8-atom ring.*

We define the maximum in-plane stress reached for each GB structure as the in-plane strength, $f_u$, and the corresponding strain as the ultimate strain $\varepsilon_u$. Similarly, the in-plane stress at the strain step before fracture and the corresponding strain was defined as fracture in-plane stress, $f_f$ and $\varepsilon_f$, respectively. These values for each silicene GB structure along with their formation energy as defined in equation 5 and in-plane stiffness are summarized in Table 2. The above-mentioned values for the same GBs on other 2D materials are also included.

*Table 2: Ultimate strain ($\varepsilon_u$), in-plane strength ($f_u$), fracture strain ($\varepsilon_f$), fracture in-plane stress ($f_f$), formation energy($E_f$) and in-plane stiffness ($C$) for selected grain boundaries and the two orientations of pristine silicene. The values for graphene or hexagonal boron nitride (h-BN) grain boundaries reported in previous work are also listed for comparison. Adjacent rows shaded in same color are GBs on different 2D materials but with similar atomic structures for comparison.*

| GB | $\varepsilon_u$ (%) | $f_u$ ( $N/m$) | $\varepsilon_f$(%) | $f_f$( $N/m$) | $E_f$ (eV/Å) | $C$( $N/m$) |
|---|---|---|---|---|---|---|
| Armchair Silicene | 17 | 6.999733 | | | 0 | 62.78736 |
| Armchair Silicene (14) | 18 | 7.07 | | | | 63.51 |
| Armchair Graphene (35) | | 37.74† ($\sigma$=113 GPa) | | | | |
| Armchair h-BN (36) | 18 | 28.71† ($\sigma$=87 GPa) | | | | |
| Zigzag Silicene | 18 | 5.597882 | | | 0 | 60.91112 |
| Zigzag Silicene (14) | 14 | 5.66 | | | | 60.06 |
| Zigzag Graphene (35) | | 34.07† ($\sigma$=102 GPa) | | | | |
| Zigzag h-BN (36) | 24 | 34.00† ($\sigma$=103 GPa) | | | | |
| (1,3)\|(1,3)-i2 Silicene (Class 2) | 14 | 6.191736 | 15 | 6.148374 | 0.071 | 58.90067 |
| (1,3)\|(1,3)-i2 Graphene (35) | 15 | 31.06† ($\sigma$=93 GPa) | 15 | 31.06† ($\sigma$=93 GPa) | 0.33 | - |
| (1,3)\|(1,3)-i5 Silicene (Class 2) | 14 | 6.060064 | 15 | 6.053133 | 0.019 | 59.27565 |
| (1,3)\|(1,3)-i4 Silicene (Class 2) | 13 | 5.661474 | 14 | 5.539995 | 0.11 | 56.82485 |
| (1,2)\|(1,2)-i1 Silicene (Class 1) | 14 | 5.541226 | 15 | 5.379853 | 0.086 | 59.08894 |
| (1,2)\|(1,2)-i1 Graphene (35) | 14 | 29.73† ($\sigma$=89 GPa) | 14 | 29.73† ($\sigma$=89 GPa) | 0.34 | - |
| (1,2)\|(1,2)-i1 h-BN (36) | 20 | 24.75† ($\sigma$=75 GPa) | 20 | 24.75† ($\sigma$=75 GPa) | 0.51 | - |
| (2,2)\|(1,3)-i5 Silicene (Class 3) | 17 | 4.928155 | 17 | 4.928155 | 0.101 | 54.15928 |
| (1,2)\|(1,2)-i2 Silicene (Class 3) | 10 | 4.390043 | 15 | 3.856846 | 0.166 | 47.11179 |
| (1,4)\|(1,4)-i4 Silicene (Class 4) | 12 | 4.15911 | 13 | 3.534918 | 0.125 | 46.7449 |
| (1,2)\|(1,2)-i3 Silicene (Class 1) | 11 | 4.103552 | 17 | 3.160014 | 0.104 | 48.65941 |
| (2,3)\|(2,3)-i3 Silicene (Class 6) | 16 | 4.612402 | 18 | 4.327349 | 0.147 | 46.09683 |

*† Labelled values were calculated using equation: $f = \sigma t$ based on the cited work, the corresponding stress $\sigma$ and thickness $t$ in the cited work were used. $t$ equals to 0.334 nm (35) for graphene and 0.33 nm (36) for hexagonal boron-nitride.*

From Fig. 7 and Table 2, it can be seen that the in-plane stiffness, $C$, for pristine armchair and zigzag silicene orientations were 62.8 $N/m$ and 60.9 $N/m$, respectively. In addition, the in-plane strength for armchair and zigzag silicene was 7.00 $N/m$ and 5.60 $N/m$, respectively. The in-plane stiffness are in good agreement with the previously reported values of 63.51 $N/m$ and 60.06 $N/m$ for armchair and zigzag directions, respectively(14). It can be seen from Table 2 that, the in-plane strength and fracture in-plane stress of silicene GBs with same translation vectors are much lower than those of h-BN and graphene. Unlike black phosphorene which has in-plane stiffness equals to 99.38 $N/m$ and 14.62 $N/m$ (38) (values calculated using $C = Et$, with $E$ equals to 179.07 GPa and 26.35 GPa and $t$ equals to 5.55 Å as reported in the cited work) for zigzag and armchair direction, respectively, pristine silicene does not have highly anisotropic mechanical strength.

It can be seen that the structures based on the pentagon-heptagon pairs (class (1) and (2)) had much larger in-plane strength compared to other classes, except for (1, 2)|(1, 2)-$i3$. The in-plane strength of structures (1, 3)|(1, 3)-i2, (1, 3)|(1, 3)-i5 and (1, 3)|(1, 3)-i4, which belong to class (2), have an in-plane strength of 6.19 N/m, 6.06 N/m and 5.66 N/m, respectively. These values are larger than the in-plane strength of the pristine sheet along the zigzag direction. The structure (1, 2)|(1, 2)-i1, which belongs to class (1) has an in-plane strength of 5.54 N/m, which is comparable to that of the zigzag orientation of the pristine sheet. However, for the GB structure (1, 2)|(1, 2)-$i3$, which is also based on the pentagon-heptagon pairs, the in-plane strength is much lower. This deterioration of the mechanical strength is due to the atomic feature of an adatom is located at the hollow site of the heptagon. The effect of this adatom is discussed later in this section. For in-plane stiffness, $C$, a similar trend was observed as the in-plane strength. It can be seen that the majority of GB structures belonging to classes (1) and (2) had in-plane stiffness values ranging from 56.8 N/m to 59.3 N/m, which is higher than other GB structures and comparable to a zigzag orientation of pristine silicene. By comparing $C$ and $f_u$ for (1, 3)|(1, 3)-$i2$ and (1, 3)|(1, 3)-$i4$, we conclude that the adatom at the hollow site of a hexagon near the GB weakens the GB structure.

As shown in Fig. 8(a), both in-plane strength and the in-plane stiffness correlate with formation energy and decreases with increasing formation energy. This is similar to flat graphene GBs that do not have any inflection angles (35). However, comparing the in-plane stiffness and strength between silicene and graphene GBs, as shown in Table 2, it is worth noting that graphene and h-BN, either pristine or with GBs, have much higher values for in-plane strength than silicene.

As mentioned, the mechanical behavior of SW defect GBs, except (1, 2)|(1, 2)-$i3$, do not significantly affect the mechanical strength of the silicene sheet. Conversely, other defects dramatically lower the stiffness and the in-plane strength. The structure (1, 2)|(1, 2)-$i3$, as shown in Fig. 3(e), has SW defects in the GB region, yet has a much lower in-plane stiffness (48.7 $N/m$) as compared to the other structures from classes (1) and (2). We attribute this anomaly to the prestrain of the bonds formed by the adatom and the heptagon as indicated in Fig. 8(b).

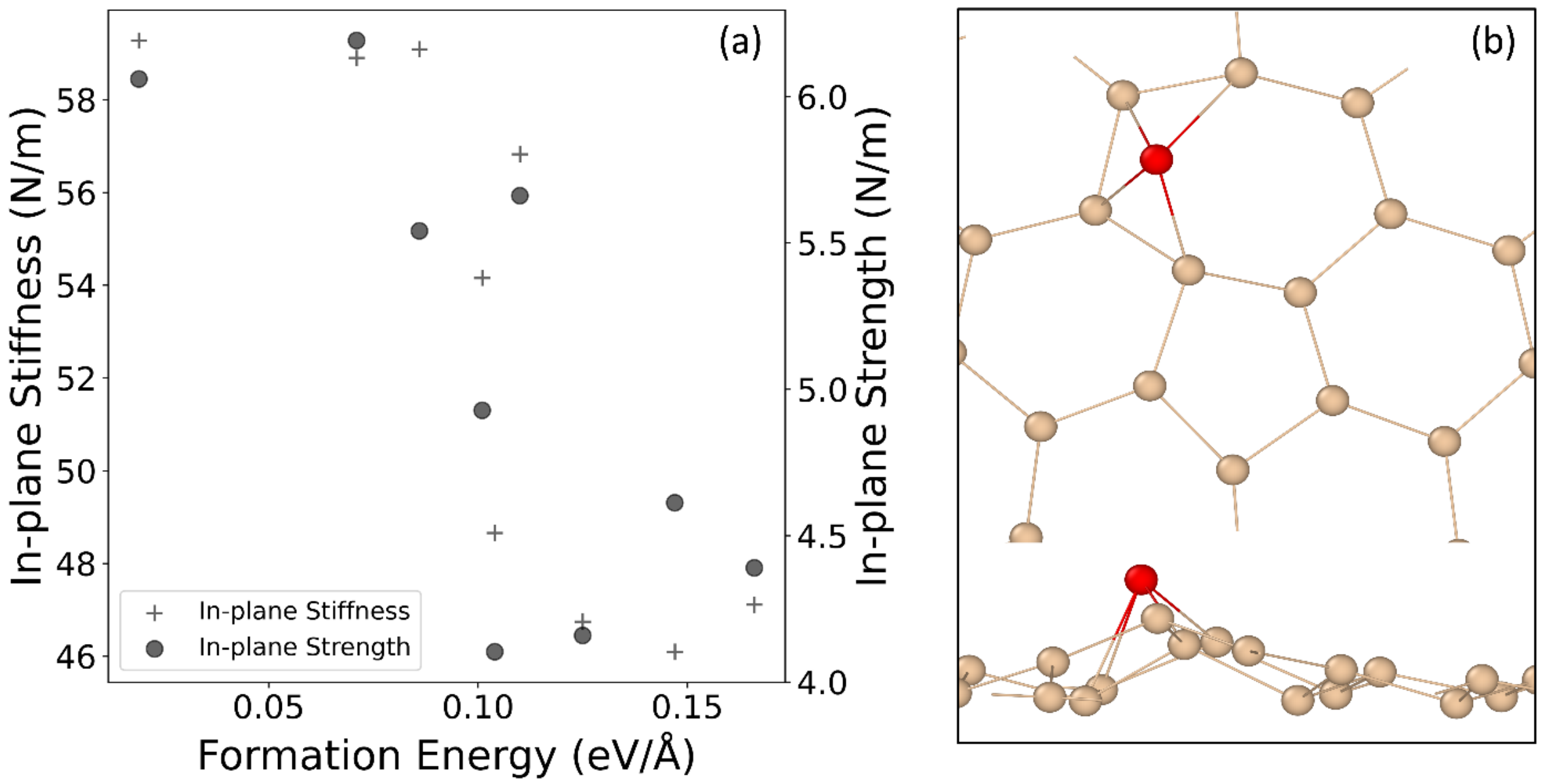

*Figure 8: (a) In-plane stiffness and in-plane strength as a function of formation energy of the structures. (b) Top and side view of the structure (1, 2)|(1, 2)-i3 with an adatom at the hollow site of the heptagon and the bonds highlighted in red. A large prestrain is observed for the adatom bonds.*

The average bond length for adatom bonds in a hollow site, as shown in Fig. 8(b) (highlighted in red), is 2.61 Å and is much larger than that of pristine silicene, which has an average bond length of 2.28 Å. This indicates that adatom bonds at a hollow site possess a larger prestrain, and therefore lead to a lower in-plane stiffness. Also, as can be seen from the video file for the strained (1, 2)|(1, 2)-$i3$ structure in the supporting information, when the stepwise strain is applied to the sheet, the adatom moves to within the 2D plane and forms a four-atom ring.

### 3.3. Fracture Behavior

In addition to the differences for in-plane stiffness and in-plane strength, the fracture behavior for the GB structures based on pentagon-heptagon pairs was also very different from other GBs. We define the GB structures that have the same translation vector combinations but with different topological structures at the GB region as different phases, and unlike h-BN (36), for which most of the GB structures show brittle fracture characteristics, some of the silicene GBs undergo phase changes before fracture. The majority of GB structures that belong to classes (1) and (2), except for (1, 2)|(1, 2)-$i3$ and (1, 3)|(1, 3)-$i4$, did not undergo any phase changes before the brittle fracture. For (1, 3)|(1, 3)-$i4$, as shown in Fig. 4(b), the adatom at the hollow site of the hexagon migrates to the top site of a neighboring silicene atom during the strain steps, as shown in the strain video in the supporting information. However, no in-plane stress decrease is observed during the straining process and the fracture behavior for (1, 3)|(1, 3)-$i4$ is also consistent with majority of the GBs based on the pentagon-heptagon pairs where fracture tends to initiate from the atomic bond shared by a heptagon and hexagon ring (7-6 bond). This can be seen from the videos showing straining and fracture in the supporting information for (1, 3)|(1, 3)-$i2$, (1,3)|(1,3)-i5, (1, 3)|(1, 3)-$i4$ and (1, 2)|(1, 2)-$i1$ structures. The same fracture mechanism has also been observed for graphene (35), and just like graphene, this behavior can be attributed to the prestrain of the 7-6 bonds. In contrast, for the (1, 2)|(1, 2)-$i3$ structure, the fracture behavior is altered by the adatom at the hollow site of the SW defects. During the straining process, the original structure of (1, 2)|(1, 2)-$i3$ first transforms to periodically repeating heptagon and pentagon rings separated by a four-atom ring that is generated when the adatom moves to within the 2D plane, as shown in Fig. 9(a). Then, the fracture initiates from the bond that is shared by the hexagon and the four-atom ring (6-4 bond). From there, fracture propagates to the 7-6 bond as mentioned above. By comparing the failure behavior between the various GB structures belonging to classes (1) and (2), we conclude that the adatom on a top site and a hollow site of a hexagon does not affect the failure behavior of the structure nor the mechanical strength. In contrast, an adatom within a hollow site of a heptagon will significantly impact the stiffness and modify the failure behavior. One may notice that this result seems to contradict previous results for the formation energy, where an adatom within a hollow site of the hexagon have higher formation energy due to the smaller area of hexagon and larger strain. However, it worth noticing that the adatom within the hollow site of the hexagon migrates to the top site when a strain is applied. We believe this phase change due to strain is the reason why a SW GB structure with an adatom within a hexagon has comparable mechanical strength and similar fracture behavior to the GB structures based on SW defects.

Additionally, the (2, 2)|(1, 3)-$i5$ and (2, 3)|(2, 3)-$i3$ structures undergo a phase transition when strain is applied, as shown in video files in supporting information and Fig.9. The structure (2, 2)|(1, 3)-$i5$, as shown in Fig. 4(f), transformed to a predominantly SW defect GB structure when 6% strain was applied to the sheet, as shown in Fig. 9(b). This phase transition corresponds to the sudden gradient change at the corresponding strain step for the curve, labeled with purple octagons as shown in Fig. 7. After the phase transition of (2, 2)|(1, 3)-$i5$, it is worth noting that the in-plane stress continues to increase with increasing strain until the ultimate strain is reached (17%). The increase of the in-plane stress after the phase transition is due to the newly formed pentagon-heptagon pairs in GB region, and results in the ultimate strain being much larger than for other

structures that do not belong to class (1) or (2). For the structure $(2, 3)|(2, 3)\text{-}i3$, as shown in Fig. 5(g), the triangular rings formed near octagons transform into four-atom rings as the silicene atoms relocate to within the plane when 6% strain is applied to the sheet, as shown in Fig. 9(c). The in-plane stress continues to increase with increasing strain after this phase transition. The gradient change at 12% applied strain corresponds to the fracture of the bond that is shared by the octagon and 5-atom ring (8-5 bond) and failure occurred at 19% strain. Structures $(1, 2)|(1, 2)\text{-}i2$, as shown in Fig. 4(d), and $(1, 4)|(1, 4)\text{-}i4$, in Fig. 5(a), did not experience any phase transition before failure occurred. The gradient changes on the curve shown in Fig. 7 correspond to a series of bond fractures during the straining process. In summary, all the GBs experiencing phase transition during the straining process have out-of-plane atoms and the out-of-plane atoms translate to within the plane when a large enough strain is applied to the sheet.

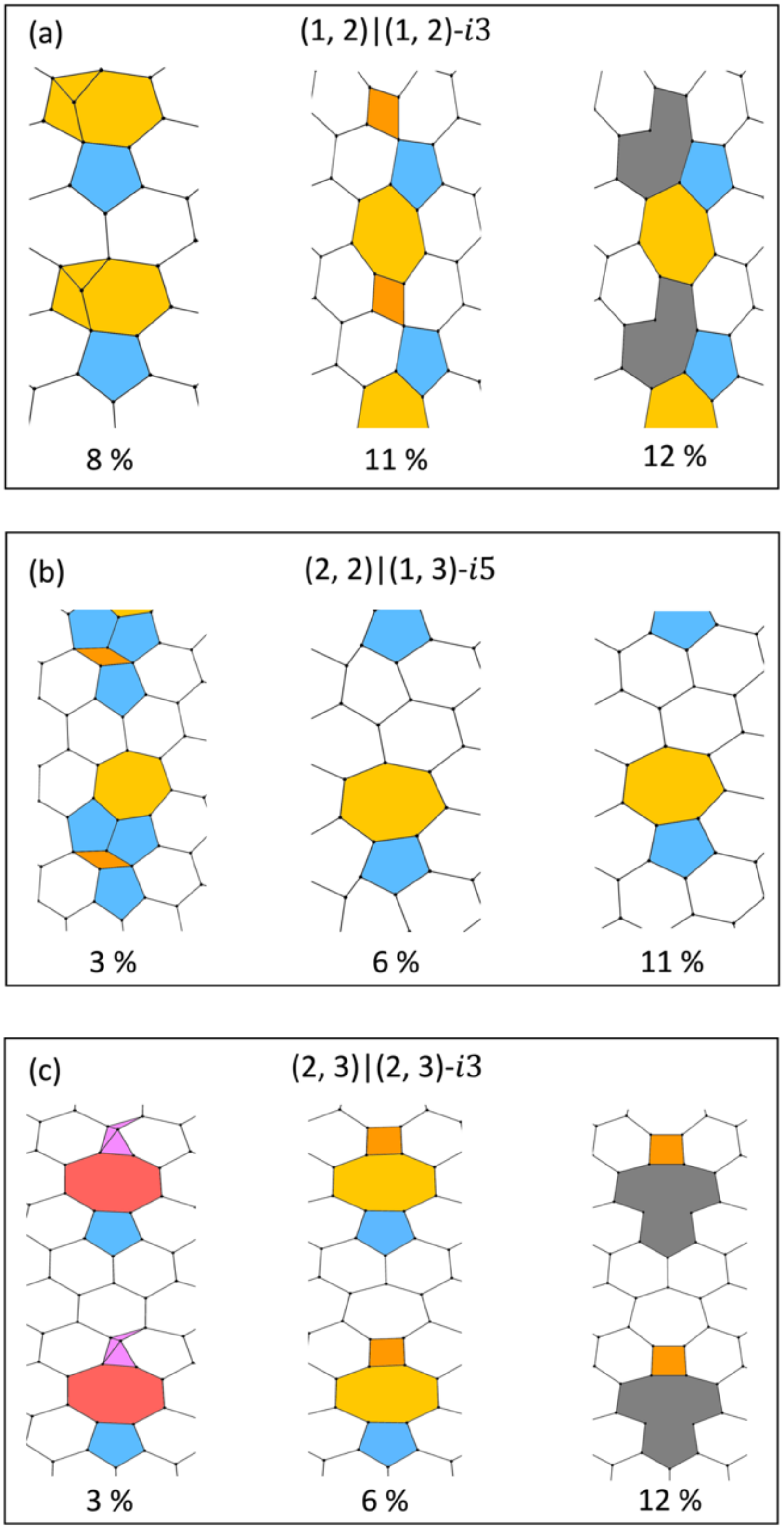

*Figure 9: Grain boundary structure (a) (1, 2)|(1, 2)-i3, (b) (2, 2)|(1, 3)-i5 and (c) (2, 3)|(2, 3)-i3 under various strain.*

# 4. Conclusion

In summary, using evolutionary algorithm with empirical potential, we have performed a comprehensive search of energetically favorable GB structures for silicene. Several previously unreported novel GB structures were predicted. Multiple atomic GB structures such as adatoms, four-atom rings and additional pentagons for SW defects were found that have not been seen in the GBs for graphene and h-BN. Further theoretical study shows that in general, the formation energy of silicene is lower than graphene and h-BN, and the location of Si adatoms within the silicene GB region may increase or dramatically lower the formation energy depending on where it is located and chemical bonds it forms with neighboring atoms. We also found that the GBs based on pentagon-heptagon pairs generally have lower formation energy, and presence of four-atom rings lower the formation energy of GBs by relieving the strain for atomic structures.

Using first-principles calculations, a uniaxial strain perpendicular to the GBs was also applied to the structures to study the mechanical properties of GBs. The GBs based on the SW defects generally have comparable or even higher in-plane strength and in-plane stiffness than the pristine sheet. The fracture behavior for these GBs were also consistent with previously reported SW GBs in graphene with the exception of when an adatom is located within the hollow site of a heptagon. In general, we found the in-plane stiffness and in-plane strength of GBs decreases with an increasing formation energy of GB. The failure mechanism of the GB with heptagon hollow site adatoms were also analyzed along with other GBs that have lower in-plane strength. We summarize that the lower mechanical strength of this heptagon adatom GB can be attributed to a prestrain from the atomic bonds. All the GBs experiencing phase transition during the straining process have out-of-plane atoms and the out-of-plane atoms translate to within the plane when a large enough strain is applied to the sheet. In summary, this study provides insight on new silicene GB structures and their corresponding mechanical behavior, which shall guide the future research work and applications of silicene.

# Supporting Information

The video files showing the strain steps of the GB structures are available in the data repository:

https://figshare.com/s/24f96ed271e437714ebf

# Acknowledgements

The use of the Center for Nanoscale Materials, an Office of Science user facility, was supported by the U.S. Department of Energy, Office of Science, Office of Basic Energy Sciences, under Contract No. DE-AC02-06CH11357. This research used resources of the National Energy Research Scientific Computing Center, which was supported by the Office of Science of the U.S. Department of Energy under Contract No. DE-AC02-05CH11231. The authors would like to acknowledge the support from the Argonne LDRD and UIC faculty start-up funds. This material is based upon work supported by the U.S. Department of Energy, Office of Science, Office of Basic Energy Sciences Data, Artificial Intelligence, and Machine Learning at DOE Scientific User Facilities program under Award No. 34532.